# Optimization of heterogeneously integrated InP-Si on-chip photonic components

**Paweł Mrowiński,**[1*] **Paweł Holewa,**[1,3] **Aurimas Sakanas**[2], **Grzegorz Sęk,**[1] **Elizaveta Semenova,**[2,3] **and Marcin Syperek**[1]

[1]*Department of Experimental Physics, Faculty of Fundamental Problems of Technology, Wrocław University of Science and Technology, Wybrzeże Wyspiańskiego 27, 50-370 Wrocław, Poland*
[2]*DTU Electro, Technical University of Denmark, Kongens Lyngby 2800, Denmark*
[3]*NanoPhoton-Center for Nanophotonics, Technical University of Denmark, 2800 Kongens Lyngby, Denmark*
*\*pawel.mrowinski@pwr.edu.pl*

**Abstract:** We demonstrate comprehensive numerical studies on a hybrid III-V/Si-based waveguide system, serving as a platform for efficient light coupling between an integrated III-V quantum dot emitter to an on-chip quantum photonic integrated circuit defined on a silicon substrate. We propose a platform consisting of a hybrid InP/Si waveguide and an InP-embedded InAs quantum dot, emitting at the telecom C-band near 1550 nm. The platform can be fabricated using existing semiconductor processing technologies. Our numerical studies reveal nearly 87% of the optical field transfer efficiency between geometrically-optimized InP/Si and Si waveguides, considering propagating field along a tapered geometry. The coupling efficiency of a directional dipole emission to the hybrid InP/Si waveguide is evaluated to ~38%, which results in more than 33% of the total on-chip optical field transfer efficiency from the dipole to the Si waveguide. We also consider the off-chip outcoupling efficiency of the propagating photon field along the Si waveguide by examining the normal to the chip plane and in-plain outcoupling configurations. In the former case, the outcoupling amounts to ~26% when using the circular Bragg grating outcoupler design. In the latter case, the efficiency reaches up to 8%. Finally, we conclude that the conceptual device's performance is weakly susceptible to the transferred photon wavelength, offering a broadband operation within the 1.5-1.6 μm spectral range.

## 1. Introduction

Quantum photonic integrated circuits (qPICs) are functionalized based on the quantum nature of photons [1–3]. They can find applications in new fascinating fields related to quantum information processing and computing [4,5], quantum communication [6], or quantum metrology [6]. As in classical photonic integrated circuits (PICs), chip-scale integration is a crucial factor offering miniaturization and thus a small system footprint, scalability, mechanical stability, low energy as well as material consumption, and low manufacturing costs [1,2,7,8]. To achieve full application capabilities, the qPIC requires integrating various passive and active optical components within a single wafer-scale platform allowing for non-classical light generation, photon routing, control over the whole qPIC elements, and detection at a single photon level, or light outcoupling to the external photonic environment [1,3,7,8]. Many existing material approaches are currently considered for developing a genuinely functional qPIC, including major platforms like silicon-on-insulator (SOI), silica-on-insulator ($SiO_2OI$), silicon nitride ($Si_3N_4$), silicon oxynitride, aluminum nitride (AlN), silicon carbide (SiC), lithium niobate (LN), diamond (C), polymers, tantalum pentoxide ($Ta_2O_5$), gallium arsenide (GaAs), or indium phosphide (InP) [3,7,8]. Nevertheless, within all the mentioned, monolithic fabrication of a functional qPIC-based device is currently very challenging or impossible, routing towards hybridization of many different materials [1].

Silicon-on-insulator is a very attractive major material platform that can be used for the qPICs integration [9–11]. It has been successfully utilized in classical PICs, offering advantages such as compatibility with mature complementary metal-oxide-semiconductor (CMOS) technologies, thus allowing for electronic control over the qPICs elements [12], providing high index contrast necessary for the light routing [13] and assuring low-cost, maturity, and high processing-yield [9–11]. Within the Si platform, one can fabricate various monolithic components, including splitters [14], crossings, phase shifters, filters, electro-optical switches, waveguides [15], multiplexers [16], micro-ring resonators [17] and photon detectors [18]. Notably, the silicon platform offers ultralow losses for light routing within a chip [15,19] supported by a high refractive index [13]. For the scalability of quantum photonic devices, low photon losses within a chip are of primary importance, being the scalability limiting factor [1,2]. However, the main hurdle is the lack of a compatible non-classical photon source. It seems that, as in the case of other material platforms, hybridization of a non-classical photon generator with the Si platform might be a challenging but necessary solution [20–22].

One of the attractive solutions would be a heterogeneous integration of a III-V semiconductor quantum dot (QD) with a Si-based platform [2,20,22]. The challenge of this approach is that photons must be effectively distributed within both materials of different refractive indices [13,23]. Therefore, an efficient light field coupling must be provided between the QD, its environment, and the Si photonic platform.

It has already been demonstrated that a single InAs/GaAs QD placed in the host GaAs waveguide (WG) can be effectively integrated with a $Si_3N_4$ WG through a direct wafer bonding scheme [24]. The hybridization scheme has a significant potential for scalability, which has already been proven for the application to classical PICs [25]. The efficient light coupling of 90% has been achieved between GaAs and $Si_3N_4$ WGs through the WGs tapering [24]. This integrated GaAs/$Si_3N_4$ system also demonstrated on-demand indistinguishable photons distributed over the $Si_3N_4$ WG [26]. Despite the successful presentation of the proof-of-concept, the hybrid GaAs/$Si_3N_4$ material system suffers from limitations related to the useful spectral range of the transmitted photons. For example, having a single photon source compatible with the telecom industry standards exploiting 1.3 μm or 1.55 μm optical transmission windows of silica fibers eliminate the problem of frequency converters that can be an essential source of losses for a hybrid III-V/Si qPIC used in distant quantum communication or distributed quantum computation applications [5]. While InAs/GaAs QDs can be forced to emit at near-infrared by utilizing strain tuning [27–31], InAs/InP QDs [22,32–35] can naturally emit photons in the mentioned spectral ranges. A particular interest is paid to the 1.55 μm transmission channel. It promises low photon losses and single-mode propagation due to a high refractive index contrast [36] and thus strong field confinement providing effective transmission in Si WGs and silica fibers at large distances.

In this work, we propose hybridizing an InAs/InP QD embedded in an InP WG with the Si WG fabricated on the SOI platform for future Si-based qPICs. We provide a process flow chart which is fully feasible and compatible with today's technology by utilizing a direct wafer bonding approach. The main problem tackled in this work is how to efficiently transfer photons originating from an InAs/InP QD emitting at 1.55 μm from the host InP WG to the Si WG. Our studies are realized numerically employing the finite-difference time-domain (FDTD) method implemented in Lumerical – a commercially available FDTD solver [37]. Lumerical has proved its capabilities in optimizing of various nanophotonic and optoelectronic devices [38–41], and is also recognized as a potent tool for studying light propagation in nanophotonic structures.

In the first part of the work, we investigate the transmission of the tapered hybridized InP/Si WG section, as the vision is to provide the highest coupling of InAs/InP QD-confined dipole emission to Si WG via maximized light transmission through the tapered geometry. It is realized by scanning an extensive parameter space of the taper geometry and then simulating the multimode propagation along the number of taper structures. Next, we evaluate a dipole, directional emission coupling to guided modes confined in a specified hybrid InP/Si WG using a centered dipole position, and as a function of cross-sectional displacements. Finally, we study the possibility of emission outcoupling from the Si WG to free space that simulates an interlink between the qPIC and the off-chip transmission channel, e.g., an optical fiber [42]. Each component's transmission/coupling efficiency is examined with respect to the transmitted photon wavelength in the range of 1.5-1.6 μm, which is important for the device tolerance related to the imperfection of the component's fabrication process. All of the components communicating on-chip could compose an envisioned nanophotonic circuit for quantum information processing, as schematically demonstrated in Fig. 1.

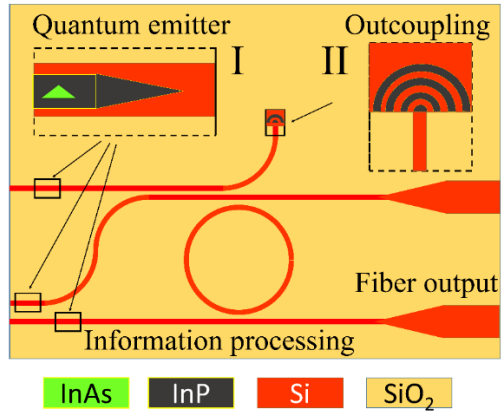

Fig. 1. Concept of a nanophotonic circuit for quantum information processing utilizing InAs/InP quantum dot emitters embedded in hybrid InP/Si waveguides on a common SOI platform. A typical nanophotonic chip may consist of photon sources, couplers, splitters, ring resonators, and light outcoupling systems off the PIC towards in-plain or out-of-plane directions of the chip.

**2. Method and model structure**

The geometry of InP and Si WGs and a hybrid WG are modeled within the Lumerical 3D CAD environment. Although the model is rather conceptual, our main concern is to find a realistic and a consistent model with the well-established fabrication and processing steps. There are a few possible approaches to realize these structures, including (i) *in-situ* lithography [43,44], (ii) transfer printing [45,46], or the (iii) pick-and-place method [47]. For each of these techniques, at least one challenge can be identified: (i) the complex equipment needed, (ii) the low fabrication precision, and (iii) the low process yield, respectively. On the other hand, direct wafer bonding followed by electron beam lithography (EBL) was successfully applied to fabricate GaAs/$Si_3N_4$ waveguides [24]. The benefits of the bonding approach are a relatively low complexity that leads to much better scalability and high fabrication accuracy (<50 nm [48], as compared with ~1 μm for transfer printing [49]). Notably, the results of the presented geometry optimization are independent of the fabrication method, however, to prove the usability of our approach, we give the details of the proposed fabrication based on direct wafer bonding. It requires the integration of an epitaxially grown InP wafer containing an array of self-assembled InAs quantum dots with an SOI wafer and subsequent WGs processing. The processing would start with patterning Si waveguides and alignment marks in the top layer of the SOI wafer. The next step would be plasma-activated direct bonding of InP to the patterned SOI chip [50], followed by removal of the InP substrate. Then, uncovered alignment marks previously defined in the Si layer would be used as a reference for deterministic patterning of InP WGs with a QD emitter on top of Si WGs. The final pattern of hybrid WG would then be transferred to the InP layer by dry etching (using the interferometry-based metrology to control the etch rate or deposition of a thin oxide layer as an etch stop). The overview of the fabrication steps is demonstrated schematically in Fig. 2.

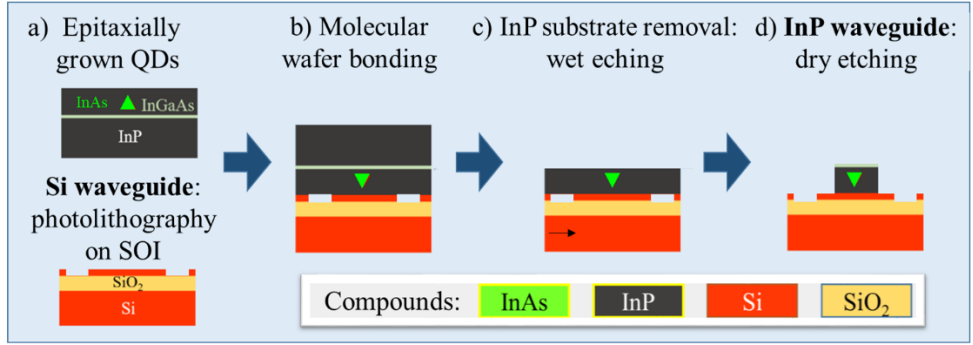

Fig. 2 The proposed fabrication workflow to realize a quantum photonic integrated circuit with heterogeneously integrated InAs/InP quantum dot system on a silicon-on-insulator (SOI) platform: a) preparation of InP wafer with quantum dots and SOI wafer with a define Si waveguide, b) molecular wafer bonding process of flipped InP wafer, c) wet chemical etching process to remove InP substrate, and d) dry etching to define InP waveguide upon Si waveguide.

We find that the main requirement for the hybrid InP/Si WG geometry, enabling confinement of the fundamental mode in the InP part, is that the Si part must be of a smaller width than the InP part. However, we expect the structure to be mechanically unstable, so the subsequent results would be considered for the opposite constraints of our conceptual design, i.e. a narrower and higher InP part is placed on top of a broader and thin Si part. The proposed fabrication method allows for achieving the hybrid InP/Si WG, as depicted in Fig. 3. It consists of three elements that can be used to develop a simple qPIC demonstrator operating at a single photon level using an embedded InAs/InP quantum emitter. The first element is a hybrid waveguide structure containing an InP WG (with a single InAs QD embedded) on top of a Si WG of different widths. The second element is a tapered structure in which both the InP and Si WGs have tapered geometries. In the case of the InP WG, the taper shrinks along the propagation direction of the mode, whereas the inverted taper geometry is used for the Si WG. Here, we consider only the case where the taper length of the InP WG is equal to or smaller than the taper length of the Si WG. Both tapers have a linear dependence on the width along the waveguide direction. The third component is the Si WG section, which is an extension of the Si WG taper end without the InP WG on top. This component is terminated by the outcoupler ring structure, which can be used to scatter the transferred light to the off-chip detection system, normal to the structure plane. Optionally, the side detection system can be used for the cleaved facet of the sample. In both cases, photon outcoupling can be realized using a microscope objective or a lensed optical fiber to enable high collection efficiencies.

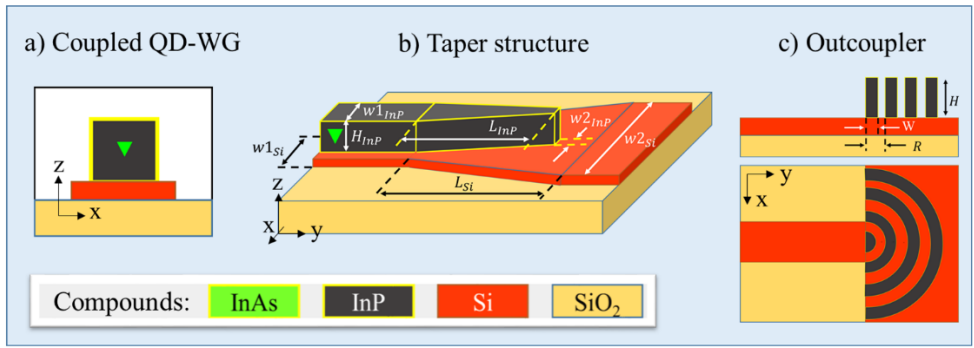

Fig. 3. Elements of the integrated quantum photonic circuit with the integrated quantum dot in a hybrid InP-Si waveguide system: a) quantum dot coupled in waveguide, b) linear taper structure, and c) grating outcoupler. All of these elements are feasible by using the proposed fabrication method in Fig. 2.

The essential material parameters for the simulations are refractive indices: $n_{InP}$ = 3.167, $n_{Si}$ = 3.475 and $n_{SiO2}$ = 1.44, which are taken for the 1550 nm photon wavelength. The size of the simulation domain in FDTD is defined as a 3x40x1 μm$^3$ box in the x-y-z Cartesian space and a mesh cell size is defined as 40x40x40 nm$^3$ (see Supp. Inf. for more details about FDTD settings). The height of the InP (Si) WG is fixed at 0.58 (0.22) μm, which gives the total height of 0.8 μm for the hybrid WG structure. In the case of Si WG, the height is determined by one of the commercially available SOI wafers. In the case of InP WG, the exact value is set after analysis of the preliminary calculations. The preliminary studies were performed for a QD embedded in the center of a 1 μm-wide InP part of the hybrid WG on top of a 1.5 μm-wide Si part, using a fully sharp, 20 μm-long taper structure. The outcoupler had a grating period R = [$\lambda_0$· (1−1/2)/$n_{InP}$ + $\lambda_0$/2]. The geometries of the taper and outcoupler were not optimized here. Varying the height of InP WG in the range of 0.4 μm to 0.8 μm, we found that for the value of 0.58 μm, one can achieve the best vertical outcoupling, reaching roughly 6%. An initial test was performed for the emission wavelength of 1.55 μm. We should note here that by using the optimized system, as it is described in Results section 3 below, and performing similar calculations in dependence on the InP height, we found that for the InP height of 0.58 μm, again, one can get the maximum efficiency for the vertical outcoupling (see Supp. Inf. and the related Figures S1-S3 for more details).

Although calculations are performed within a broad spectral range of 1.5 μm to 1.6 μm of photon wavelengths, the geometry optimization offering the best InP to Si WG coupling is tuned to the 1.55 μm wavelength.

## 3. Results

### 3.1 Light transmission from the InP to Si waveguide

In this section, we study the transmission efficiency of a propagating light field from the hybrid waveguide section, through the taper region, to the Si WG section. To simulate the light field generated by a single InAs quantum emitter embedded in the InP WG, we use a Gaussian beam source (waist radius of 0.50-0.55 μm is used depending on w1$_{InP}$) providing directional propagation of the field along the waveguide. The directional propagation of the Gaussian beam along the waveguide simplifies and generalizes the problem of finding efficient taper structure. First, the source excites a multimode propagation similarly to a random set of electric dipoles, which is related to a position uncertainty inside the InP WG expected for random in-plane distribution of self-assembled QDs in the InAs layer, and to the vertical shift of the QD layer determined during the growth process. Second, the directionality of the Gaussian beam source allows one to evaluate the taper transmission efficiency T for half of symmetric bidirectional dipole coupling. Third, due to the not preferred order of refractive index contrast, i.e. $n_{InP}$ < $n_{Si}$, the Gaussian beam is located in InP circumventing excitation of the fundamental mode confined in Si, favouring higher-order modes relevant for the QDs embedded in InP. In this approach a single-shot simulation for a given geometry of the taper structure provides reliable results instead of time and resources consuming single-dipole modelling, and then by a parameter scan of the taper geometry we can search for optimized system with a maximum efficiency of the light transfer from hybrid WG to Si WG in a more efficient way.

The light confined in the Si WG after the coupler section (tapered area) is collected at the end of the last section of the structure (the end of Si WG). Therefore, the respective light transmission can be calculated as $T = P_{Si\,WG}/P_{source}$, where $P_{Si\,WG}$ is the optical power collected at the end of the Si WG and $P_{source}$ is the optical power injected into the hybrid InP/Si WG. According to the proposed hybrid WG geometry (see Fig.3), we evaluated the optical transmission efficiency by examining the space of the following hybrid WG parameters:

$L_{InP}$, $L_{Si}$, $w1_{InP}$, $w2_{InP}$, $w1_{Si}$ $w2_{Si}$. These parameters are assigned according to Fig. 3(b) and span the ranges given in Table 1:

| $20\ \mu m \leq L_{InP} \leq 25\ \mu m$ | $0.5 \leq w1_{InP} \leq 1.5\ \mu m$ | $1.25\ \mu m \leq w1_{Si} \leq 2.0\ \mu m$ |
|---|---|---|
| $20\ \mu m \leq L_{Si} \leq 25\ \mu m$ | $0 \leq w2_{InP} \leq 0.75\ \mu m$ | $1.75\ \mu m \leq w2_{Si} \leq 2.5\ \mu m$ |

Table 1. Range of geometrical parameters of the taper structure used in FDTD simulations.

We also impose the following conditions: $L_{InP} \leq L_{Si}$, $w1_{InP} \geq w2_{InP}$, $w1_{Si} \leq w2_{Si}$ and $w1_{InP} \leq w1_{Si}$ to fulfil the hybrid WG geometry requirements in Fig. 3(b). This constraint reduces the total number of iterations in our FDTD simulations from 2880 to 1614.

In Fig. 4(a), we demonstrate the distribution histogram of the evaluated transmission coefficient T. The statistics of the results show T above 60% for the vast majority (74.8% cases) of a relatively simple taper structure and for a coarse scan of the geometries. It indicates that the tolerance in processing such tapered WGs would be relatively high, requiring the fabrication precision on the order of a few tens of nanometers for the tapered InP and a few hundreds of nanometers for inversely tapered Si WGs, respectively. Thus, while electron-beam or deep-UV lithography should be used for fabricating InP WGs, the requirements for the fabrication of Si WGs are relaxed, and a simpler UV lithography technology could be employed.

Next, a performance as high as 85% is found. The respective geometric parameters and the transmission evaluated for each case from the coarse scan are presented in Table 2.

|     | $w2_{InP}$ [μm] | $w2_{Si}$ [μm] | $w1_{InP}$ [μm] | $w1_{Si}$ [μm] | $L_{InP}$ [μm] | $L_{Si}$ [μm] | T [%] |
|---|---|---|---|---|---|---|---|
| #1 | 0.25 | 2.00 | 1.50 | 1.50 | 25 | 25 | 86.0 |
| #2 | 0.25 | 2.00 | 1.50 | 1.75 | 25 | 25 | 85.5 |
| #3 | 0.25 | 2.25 | 1.50 | 1.50 | 25 | 25 | 86.4 |
| #4 | 0.25 | 2.25 | 1.50 | 1.75 | 25 | 25 | 85.7 |
| #5 | 0.25 | 2.50 | 1.25 | 1.75 | 25 | 25 | 85.1 |

Table 2. Range of geometrical parameters of the taper structure used in FDTD simulations.

At this stage, it is worth noting that all the best cases are obtained for the following:
(i) a non-zero InP WG taper tip: $w2_{InP} = 0.25\ \mu m$;
(ii) equal lengths of both tapers: $L_{InP} = L_{Si} = 25\ \mu m$;
(iii) when an InP WG taper width decreases by 40 nm over 1 μm, a Si WG taper width increases by 20 nm over 1 μm.

Next, we performed numerical studies for each of the 5 cases presented in Table 2 using a finer parameter space. We varied both w2$_{InP}$ and w2$_{Si}$ by ±0.05 μm, ±0.10 μm, ±0.15 μm. The distributions of the results for all 5 cases are shown in Supp. Info. (Fig. S6) and show no local extrema with a slight variation in $T$ within ±3 %. The highest transmission reaching 87.1 % is found for the taper, like in case #4, for −0.1 μm detuning of the w2$_{Si}$ parameter, as shown in Fig. 4(b). On the contrary, a maximum 5% reduction in T is expected when changing one of the parameters by +0.15 μm. In this context, a typical 40-50 nm EBL precision [48], which is better than <1 μm one expected for micro-transfer printing [49], favouring the suggested processing based on wafer bonding to realize the taper structure.

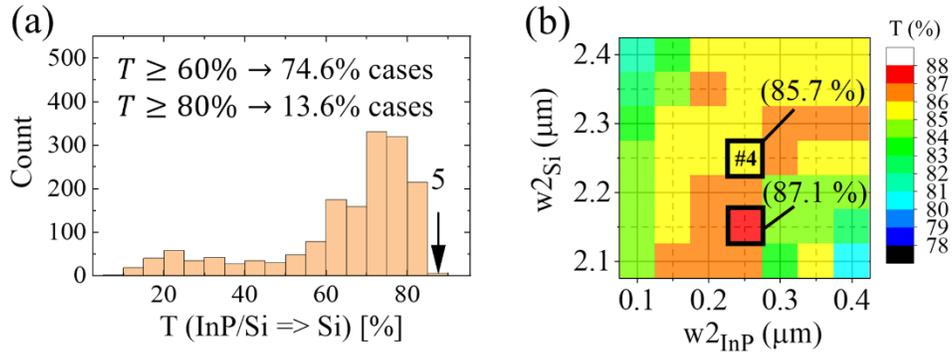

Fig. 4 (a) Histogram of the transmission coefficients T evaluated for the tapers calculated for 1614 geometries (see text for details), (b) Evaluation of taper transmission for fine tuning of taper tip widths w2InP and w2Si around geometry #4 (see Table 2) with the highest transmission found in (a).

Moreover, we observe that the maximum values in T are obtained for the longest tapers of a parameter space given in Tab. 1, therefore we performed additional simulations in dependence on tapers length from 10 to 38 μm for the most efficient taper shown in Fig. 4(b). In principle, due to the minimized field divergence along the taper, higher transmission is expected for a longer linear taper [51]. Our results presented in Fig. 5(a) are consistent with that expectation, and we observe an increase of transmission with the taper length reaching a maximum value of 89.1 % for the 36 μm long taper, however, the trend is not continuous and a further increase of the taper results in a reduced transmission. The discontinuities have a form of local minima likely related to the reflection from the taper tip. According to our above observation (i) from a coarse scan, a finite tip width is preferred to obtain the highest T, and therefore the system is susceptible to unwanted reflections. From the technological point of view, fabrication of longer tapers would also lead to fluctuations of their geometry deteriorating the field transfer to a Si-WG. Therefore, a taper length limited to about 25 μm could be more practical to obtain a highly efficient light transfer from the InP/Si to Si WG.

Last, we evaluated the taper transmission in dependence on the relative shift between the principal axes of the InP WG and the Si WG. The corresponding results that account for up to 200 nm mismatch are demonstrated in Fig. 5(b). One can see that taper transmission is reduced only by 5.3 % within the considered mismatch range. Consequently, considering the accuracy of EBL (40-50 nm) used for the proposed fabrication method, the taper transmission should still be at a very high level, above 85.8 %.

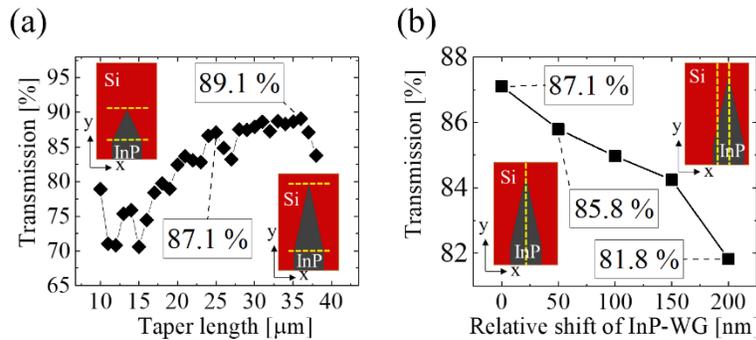

Fig. 5 Evaluation of taper transmission in dependence on (a) a taper length and (b) a relative shift between the principal axes of the InP waveguide with respect to the Si waveguide.

Furthermore, we took a closer look at the evaluated propagation of the optical field along the hybrid WG structure. Figure 6 presents the field distribution of the hybrid WG structure, which exhibits the highest transmission to the Si WG. In Fig. 6(a) we plot the Gaussian input beam in the x-z plane in the InP region. The propagation of the field along the taper region, the y-z plane (integrated over 'x' dimmension), is shown in Fig. 6(b). Finally, the field distribution at the Si WG output is depicted in Fig. 6(c). Interestingly, Fig. 6(b) shows the field oscillations between the InP and Si WG regions along the entire area of the taper. Such a field propagation character over the hybrid WG structure indicates that the taper length must be carefully adjusted to avoid reflections from the InP finite tip and to obtain a highly efficient transfer to the Si region. In addition, it suggests a multimode character of the field distribution in the tapered waveguide section.

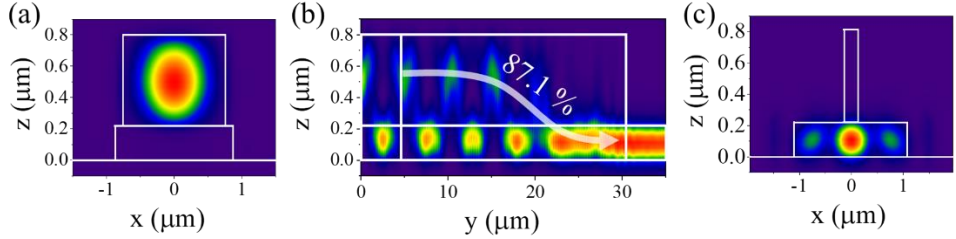

Fig. 6. Visualization of light transmission through a tapered hybrid WG structure (a) Normalized input Gaussian $|E|^2$ field centered in the InP WG, (b) normalized integrated $|E|^2$ field distribution over the 'x' axis in the tapered region, and c) output $|E|^2$ field distribution in the Si WG.

*3.2 Mode analysis and dipole coupling*

The analysis of propagating optical field modes is essential to determine the coupling between an emitting dipole and a WG. The in-plane-oriented dipole (excitation in a QD characterized by the in-plane-oriented dipole moment) is placed in the hybrid InP/Si WG section, in the center of the InP part, and the resulting in-plane linear polarization coincides with the TE-like propagating modes. The analysis of the InP/Si WG mode starts with a fixed Si WG width of 1.5 μm, scanning the InP width from 0.5 μm to 1.5 μm. The assumed 1.5 μm width of the Si WG already supports the propagation of the fundamental TE-like mode. Figure 7(a) displays the calculated real part of effective refractive indices ($n_{eff}$) plotted as a function of InP width. It shows that the mode structure evolves, and for a broader InP WG, the $n_{eff}$ increases. Interestingly, although one would expect that a higher $n_{eff}$ would be translated into a longer effective propagation because of stronger mode confinement, it turns out that this is not the case for the fundamental mode (TE-like). Increasing the InP width increases the mode transfer from Si to InP. The latter has a lower refractive index, so the excess loss, related to a propagation leakage along the WG, is getting higher. However, we note that for the higher order modes, as $n_{eff}$ increases, the excess loss decreases with increasing the InP width from 1.0 μm to 1.5 μm, as the field is distributed mode towards InP, so the leakage is effectively reduced for wider WGs. Detailed characteristics of the excess loss due to leakage for fundamental and higher-order modes are illustrated in Supp. Inf. (see Fig. S5).

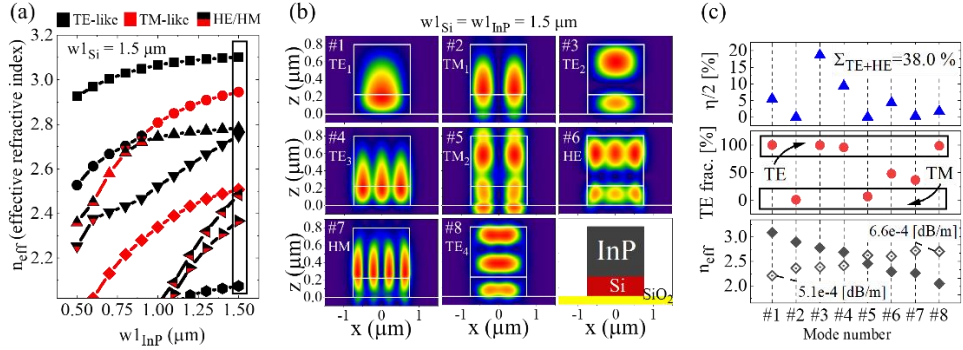

Fig. 7 (a) Real part of the effective refractive index ($n_{eff}$) for the hybrid InP/Si waveguide structure with fixed $w1_{Si}$ = 1.5 μm demonstrated as a function of $w1_{InP}$. The selected region is the case analyzed in (b) where the field distribution of the propagating modes is displayed (modes #1-#8 are top to bottom in (a)). The results shown in (c) contain the dipole (QD) coupling to the specific mode (see the article text), the TE-fraction of the field and the effective refractive index together with the excess loss.

Now, in the case of 1.5 μm-wide InP WG, for which the excess loss is minimized, we analyze the electric field distribution for the modes. Figure 7(b) shows the normalized integrated intensity for the first eight propagating modes (#1 - #8). The mode character – TE or transverse magnetic (TM), a QD (dipole) to the hybrid WG mode coupling, and excess losses are recovered, and the output results are summarized in Fig. 7(c). The mode character is evaluated by calculating the TE field fraction in the total distributed field by:

$$TE_{fraction} = \frac{\int |Ex|^2 dxdz}{\int (|Ex|^2+|Ez|^2)dxdz},$$

where {Ex,Ez} are electric field components in the WG cross-section.

The dipole-WG mode coupling is examined by simulations performed for the 35 μm-long hybrid WG structure (convergence is observed around 30 μm) with emission from the in-plane polarized dipole source embedded in the center of the InP WG (x = 0 μm; z = 0.51 μm). The light field propagates over 35 μm, and is finally collected by the monitor at y = 35 μm. Then, the overlap $\eta$ between the scattered field {**E**,**H**} and the guided mode {**E$_f$**,**H$_f$**} is calculated by using [52]:

$$\eta = \frac{Re\{\iint (E_f \times H^*)\cdot \hat{y}dS \iint (E \times H_f^*)\cdot \hat{y}dS\}}{Re\{\iint (E_f \times H_f^*)\cdot \hat{y}dS\}Re\{\iint (E \times H^*)\cdot \hat{y}dS\}}.$$

The total overlap of dipole emission is calculated as a sum for of all propagating modes. As the light emission from a point source embedded in the center of the InP WG is bidirectional, the effective dipole coupling evaluated for our system is $\eta/2$. On the other hand, applying a grating system on one side of the WG [53] or setting a dipole in a chiral position [54] could significantly enhance the directionality of emission and improve the on-chip coupling of QD emission.

Based on the evaluated TE-fraction of the modes, we conclude that in this case, we have 4 x TE-like, 2 x TM-like modes, and 2 x hybridized modes. First, we need to note that a fundamental mode (TE-like) is partially localized in both InP and Si WG sections (see Fig. 7(b)). In the set of higher order modes, one can find the modes localized more in the InP WG, like the already discussed mode #3 (also TE-like). As described in the analysis above, the dipole coupling between the InAs QD emitter inside the InP WG and the fundamental mode is less effective (~6 %) than the coupling with mode #3 (~20 %). In Fig. 7(c), we also show the losses for all the confined modes in the system presented in Fig. 7(b), which are calculated as

$$loss = -10\log_{10}\frac{P(y)|_{y=1\,[m]}}{P(y)|_{y=0\,[m]}},$$

where P(y) is the power of the field at position 'y'. We observe only a slight increase in the losses from the fundamental mode (#1) to the higher-order modes. This indicates that contributions from all the TE-like modes and HE mode (a hybrid mode in which the TE component dominates) should be considered in estimating the dipole coupling efficiency. In contrast, the coupling with the TM-like modes is less effective due to the mismatch with the in-plane linear polarization of the dipole. A quantitative study shows that the integrated overlap of all the TE-like modes including HE mode and a single QD isotropic emitter is 38.0 %, which could be twice higher for a directional emitter.

We also evaluate a possible change of the dipole coupling to relevant TE modes of the hybrid WG employing the geometry used in Fig. 7(b,c). Here, we consider the vertical displacement $\Delta z$ and the horizontal displacement $\Delta x$. The results are shown in Fig. 8. In the first case, we find that the overall coupling increases when the dipole is shifted upward along the 'z' axis of the InP WG, reaching 42.3 %. On the other hand, the displacement along the 'x' axis of the WG up to $\Delta x = 0.5$ μm reveals an oscillatory-like behaviour in coupling efficiency. A more detailed analysis of the coupling to individual modes (see Supp. Inf. Fig. S7) show that from $\Delta z = -0.1$ μm to $\Delta z = 0.1$ μm the coupling to higher order modes, i.e. #3 and #6 increases, while the coupling to fundamental mode, #1, is less effective. The minimum observed at $\Delta x = 0.25$ μm corresponds to a significantly reduced coupling to a HE mode (#6), and a reduced coupling to both fundamental modes, #1 and #3, although, it is compensated by an activated coupling of about 10% to mode #4. A further change along the axis 'x' to $\Delta x = 0.4$ μm results in a still high overall coupling of 37.5 % due to a restored coupling to HE mode (#6) and a still significant contribution of the coupling to mode #4.

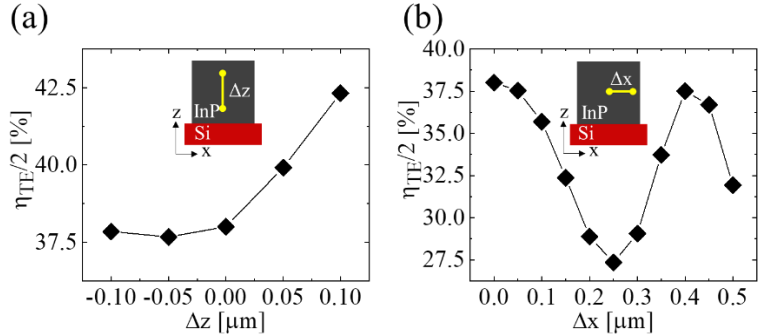

Fig. 8 Evaluated directional dipole emission coupling for a hybrid InP-Si waveguide system for a dipole displacement along (a) the vertical axis 'z' and (b) the horizontal axis 'x'. The position assigned to $\Delta z = 0$ μm and $\Delta x = 0$ μm corresponds to the center of InP WG.

*3.3 Vertical outcoupler*

Here, we examine the possibility of outcoupling the light field from the Si WG following its transfer from the hybrid tapered WG section. We note that by using a multimode hybrid WG system, the Si WG of 0.22 μm heigth itself operates likewise in a multimode manner when $w2_{Si}$ is larger than 1.2 μm. In the context of long-range on-chip information processing, it would be favourable to apply another taper that shrinks $w2_{Si}$ toward below 1.2 μm, however, a multimode character could also be beneficial in the context of quantum interference realized with the use of spatial mode beam splitter [55]. To evaluate the field transfer from the Si WG to the external detection system we use a directional Gaussian beam source localized in Si propagating towards the intersection with a grating system. The Si WG is terminated by the outcoupler based on a circular Bragg grating (CBG) geometry (see Fig.3(c)) fabricated in InP on planar Si. Another design, which could be based on CBG fabricated directly in planar Si, due to only 220 nm of thickness, for high outcoupling efficiency it would require several CBG rings defined with high precision, which makes the device processing more complex. Alternatively, the outcoupler rings could be realized in InP on top of the planar Si region during a dry etching step, which is consistent with the same process flow applied for the taper fabrication (see Fig. 2). Therefore, the InP-made CBG height $H$ in the following analysis is also set to 0.58 μm. The outcoupler design is based on half-CBG to modulate the refractive index with a radial period $R$, defined as a sum of the InP material radial width $W$ and the width of the etched part, as schematically presented in Fig. 3(c). In this approach, we look for the optimal geometry focusing on the highest collection efficiency along the normal direction to the sample surface at $\lambda_0 = 1.55$ μm wavelength, within the numerical aperture of NA = 0.65, characteristic for typical microscope objectives used in experimental setups. Therefore, the outcoupling efficiency (OE) is defined as OE = $P_{NA0.65}/P_{SiWG}$, where $P_{NA0.65}$ is the power collected within a given numerical aperture and $P_{SiWG}$ is the power transferred along the Si WG.

First, considering terminated Si WG and the beginning of the InP/Si outcoupler section, we evaluated high modes matching between fundamental modes on the level of 60%, which, at first, leads to high transmission between sections, secondly, to efficient photon field transfer from the Si WG to the InP/Si grating. The reason is mainly related to a mismatch in refractive indices between InP and Si ($n_{InP} < n_{Si}$), which leads to a strong localization of the mode in the Si of the outcoupler section. We note that contribution of higher order modes coupling in the intersection bewteen the Si WG and CBG is negligible (see also Supp. Inf. Sec. VII. Fig. S8); however, by using different SOI wafers the transfer to InP via higher order modes could possibly be enhanced.

Next, optimal collection efficiency along the normal direction to the sample plane is expected if the Bragg condition is fulfilled, namely, when the grating period R is related to the propagating photon wavelength $\lambda_0$: $n \cdot 2\pi/R = 2\pi \cdot n_{eff}/\lambda_0$, where n = 0, 1, …, thus $R = \lambda_0/n_{eff}$. Therefore, we need to reformulate the Bragg condition using $n_{eff}$, which leads to $R = \lambda_0 \cdot (1-1/m)/n_{eff} + \lambda_0/m$, where 'm' gets a value of 2 or 4. The first term is assigned to the InP radial width (W) and the second term to the etched region. In this framework, we obtained OE of roughly 1 % at 1.55 μm for both m = 2 and m = 4, suggesting that in the case of 3D scattered field simulations, this approximation is far from optimal. Subsequently, we redefined the grating period to $R = [\lambda_0 \cdot (1-1/m)/n_{eff} + \lambda_0/m] \cdot S$, where the additional parameter S is used to scale the size of the grating system. In the case of m = 2, we performed simulations using S in the range of 0.8 – 1.8. We found that the maximum value of outcoupling efficiency is roughly 11 %, for S = 1.75. Next, we followed the simulation procedure for m = 4 and S spanned the range of 0.6 – 2.2. The obtained results of OE as a function of S in a wavelength range of 1.5 – 1.6 μm are demonstrated in Fig. 9(a). The maximum OE at 1.55 μm amounts to 26 %. It is obtained for S = 1.925, corresponding to the ring period $R \cong 1.453$ μm. Interestingly, OE > 20 % is obtained for R roughly ranging from 1.396 μm to 1.566 μm, demonstrating that the tolerance in fabrication accuracy of such a CBG outcoupler is approximately 85 nm. In addition, Fig. 9(b) shows the calculated optical far-field distribution in the cross-section of the outcoupler and the evaluated far-field distribution of the scattered field along the top direction. Simulations are performed for the highest outcoupling at 1.55 μm. In the cross-sectional view (bottom), one can see an effective field transfer from the Si planar structure to InP rings and the gradually decreasing field intensity along the direction determined by the waveguide structure. The projected far-field distribution is well-focused, leading to high collection efficiency with the NA = 0.65 microscope objective as well as by optical setups equipped with lower NA objectives (e.g., 0.4), for which OE amounts to 17 %, that is, a still reasonably high value.

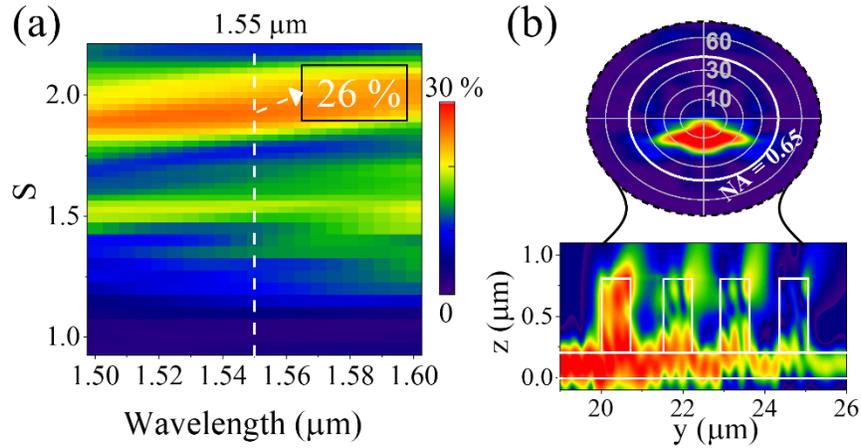

Fig. 9 a) Vertical outcoupling efficiency for a circular Bragg grating structure plotted as a function of the scaling parameter of the grating size (S) and wavelength. b) Far-field distribution of the scattered optical field collected above the outcoupler and the integrated E-field distribution in the outcoupler vertical cross-section.

*3.4 Broadband operation*

The idea of using a self-assembled single quantum dot as an emitting source in photonic integrated circuits usually suffers from some uncertainty in the QD operation wavelength due to randomness in the nanostructure size and composition. In this respect, it is required that the on-chip photonic system would have sufficient tolerance for the variability in the QD emission wavelength. The concept presented in this work is based on the elements where no cavity effects are needed. Therefore, efficient broadband operation is expected. We examined all the components introduced here regarding wavelength dependence between 1.5 μm to 1.6 μm. These components are: i) the hybrid tapered waveguide structure used for the transfer of emission from the InP to Si WG, ii) dipole coupling to TE modes confined in the hybrid WG section, and iii) the outcoupler based on a CBG for outcoupling of the field from the Si WG. Figure 10 shows the dispersion of transmission efficiencies and dipole coupling for the best taper structure (see Fig. 4(b)). Interestingly, on-chip optical field transfer maintains a very uniform overall efficiency of about 33±3 % in the entire considered spectral range, while it could be increased twice by using a directional QD-WG system by using a grating structure [53] or chiral coupling [54]. Considering the light outcoupling from the chip with the field collection optics of a high numerical aperture (NA = 0.65), one can attain a total transmission efficiency of 9±1 %, which equivalently is the on-chip transfer efficiency of bidirectional dipole emission multiplied by the outcoupling efficiency of the CBG-based outcoupler.

In addition to out-of-qPIC plane outcoupling, one can consider the in-plane outcoupling method. In this case, a Si WG fiber output port along the Si WG can be applied. With this solution, the Si WG should be terminated with an inverted taper structure, i.e. spot size converter, to enhance the in-plane directionality of outcoupling. For this type of in-plain outcoupler, we used FDTD to evaluate the incoming light scattering from the Si WG on a cleaved facet of the PIC. To simulate the collection efficiency, we used the far-field monitor placed at various distances ranging from 50 μm to 200 μm. The far-field pattern analysis using NA = 0.65 demonstrates collection efficiency ranging from 10 % to 2 %, resulting in circa 5 % of the total efficiency of the side detection. This value can be increased by utilizing the lensed optical fiber as a far-field collector aligned to the cleaved facet of the qPIC (here, the spot size converter should be used to sharpen the Si WG tip). It increases the in-plain collection efficiency to 26 % [24], resulting in roughly 8 % of the total in-plain photon field transfer efficiency in the hybrid WG system with QDs.

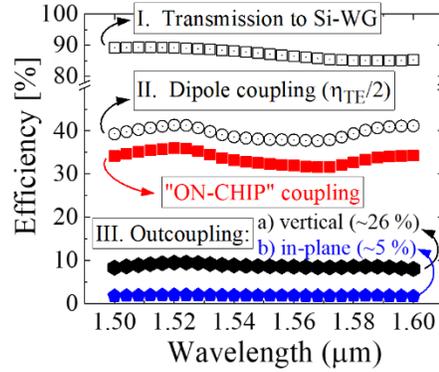

Fig. 10 Wavelength dependence of: I - transmission coefficient of the taper structure, II - dipole (QD) coupling efficiency to the TE modes and III - efficiency of the vertical outcoupler together with a combined total efficiency of about 8-10 % offered in the range of 1.5-1.6 μm.

## 4. Conclusions

In this work, we focus on the numerical investigation of the heterogeneously integrated hybrid waveguide system developed on the silicon-on-insulator platform to demonstrate its capability when used as a scalable quantum on-chip photonic integrated circuit. The structure can be fabricated by standard photo- or electron-beam lithography techniques, utilizing a bonding step to join two wafers of different materials. One of the materials can be based on the III-V compound, e.g. InP, which allows for the fabrication of high-quality quantum dot emitters, using, e.g. InAs, and operating at the 1550 nm photon wavelength. Therefore, we studied the dipole coupling in the hybrid InP/Si waveguide, the mode structure depending on the cross-sectional geometry, the linear taper structure for efficient transfer of guided light from the InP/Si hybrid system to the Si waveguide, and the possible outcoupling to interconnect with the external detection system. First, we found that the dipole coupling to the InP/Si waveguide modes amounts to 38 %. In addition, applying a reflector on one side of the waveguide [53] or employing a chiral coupling effect [54] can ensure unidirectional emission and the dipole coupling can reach up to 76 %. This number can still be significantly increased using a quantum dot-in-cavity design, where the cavity is incorporated into the InP/Si waveguide, nonetheless, for the price of narrowing the spectral bandwidth of the device. Next, we found that the photon field transfer between the InP/Si waveguide and the Si waveguide reaches 87 %. Therefore, the overall on-chip coupling, considering the light field transfer from an emitting dipole to the Si waveguide, can reach 33 %. Again, this number is highly controlled by the dipole-waveguide directional coupling and can be increased at the cost of spectral narrowing of the device's transmission bandwidth. However, we show that our conceptual device can operate in the broadband spectral range of 1.5-1.6 μm. It eliminates the problem of precise control over the quantum dot emission wavelength, which is critical for the cavity fabrication process. Finally, we investigated outcoupling efficiency realized by the circular Bragg grating-based system for the out-of-plane collection and by the spot size conversion for the in-plane collection. The results show the efficiency of the vertical outcoupling at the level of 26 % and up to 10 % of the in-plane outcoupling, both within NA of 0.65, and both showing broadband operability.

**Acknowledgements**

We acknowledge financial support from the National Science Centre (Poland) within Project No. 2020/39/D/ST5/02952 and from the Danish National Research Foundation via the

Research Centers of Excellence NanoPhoton (DNRF147). P. H. was supported by the Polish National Science Center within the Etiuda 8 scholarship (Grant No. 2020/36/T/ST5/00511) and by the European Union under the European Social Fund.

**Disclosures**

The authors declare no conflicts of interest.

See Supplement 1 for supporting content

**References**

1. A. W. Elshaari, W. Pernice, K. Srinivasan, O. Benson, and V. Zwiller, "Hybrid integrated quantum photonic circuits," Nat. Photonics **14**(5), 285–298 (2020).
2. G. Moody, V. J. Sorger, D. J. Blumenthal, P. W. Juodawlkis, W. Loh, C. Sorace-Agaskar, A. E. Jones, K. C. Balram, J. C. F. Matthews, A. Laing, M. I. Davanço, L. Chang, J. E. Bowers, N. Quack, C. Galland, I. Aharonovich, M. A. Wolff, C. Schuck, N. Sinclair, M. Lončar, T. Komljenovic, D. Weld, S. Mookherjea, S. Buckley, M. Radulaski, S. Reitzenstein, B. Pingault, B. Machielse, D. Mukhopadhyay, A. Akimov, A. Zheltikov, G. S. Agarwal, K. Srinivasan, J. Lu, H. X. Tang, W. Jiang, T. P. McKenna, A. H. Safavi-Naeini, S. Steinhauer, A. W. Elshaari, V. Zwiller, P. S. Davids, N. Martinez, M. Gehl, J. Chiaverini, K. K. Mehta, J. Romero, N. B. Lingaraju, A. M. Weiner, D. Peace, R. Cernansky, M. Lobino, E. Diamanti, L. T. Vidarte, and R. M. Camacho, "2022 Roadmap on integrated quantum photonics," J. Phys. Photonics **4**(1), 012501 (2022).
3. S. Rodt and S. Reitzenstein, "Integrated nanophotonics for the development of fully functional quantum circuits based on on-demand single-photon emitters," APL Photonics **6**(1), 010901 (2021).
4. F. Flamini, N. Spagnolo, and F. Sciarrino, "Photonic quantum information processing: a review," Reports Prog. Phys. **82**(1), 016001 (2019).
5. H. J. Kimble, "The quantum internet," Nature **453**(7198), 1023–1030 (2008).
6. R. Van Meter, *Quantum Networking*, 1st ed. (Wiley Professional, Reference & Trade, 2013).
7. E. Pelucchi, G. Fagas, I. Aharonovich, D. Englund, E. Figueroa, Q. Gong, H. Hannes, J. Liu, C.-Y. Lu, N. Matsuda, J.-W. Pan, F. Schreck, F. Sciarrino, C. Silberhorn, J. Wang, and K. D. Jöns, "The potential and global outlook of integrated photonics for quantum technologies," Nat. Rev. Phys. **4**(3), 194–208 (2022).
8. J. Wang, F. Sciarrino, A. Laing, and M. G. Thompson, "Integrated photonic quantum technologies," Nat. Photonics **14**(5), 273–284 (2020).
9. N. C. Harris, D. Bunandar, M. Pant, G. R. Steinbrecher, J. Mower, M. Prabhu, T. Baehr-Jones, M. Hochberg, and D. Englund, "Large-scale quantum photonic circuits in silicon," Nanophotonics **5**(3), 456–468 (2016).
10. S. Y. Siew, B. Li, F. Gao, H. Y. Zheng, W. Zhang, P. Guo, S. W. Xie, A. Song, B. Dong, L. W. Luo, C. Li, X. Luo, and G.-Q. Lo, "Review of Silicon Photonics Technology and Platform Development," J. Light. Technol. **39**(13), 4374–4389 (2021).
11. J. C. Adcock, J. Bao, Y. Chi, X. Chen, D. Bacco, Q. Gong, L. K. Oxenlowe, J. Wang, and Y. Ding, "Advances in Silicon Quantum Photonics," IEEE J. Sel. Top. Quantum Electron. **27**(2), 1–24 (2021).
12. J. P. G. van Dijk, E. Charbon, and F. Sebastiano, "The electronic interface for quantum processors," Microprocess. Microsyst. **66**, 90–101 (2019).
13. R. Soref and B. Bennett, "Electrooptical effects in silicon," IEEE J. Quantum Electron. **23**(1), 123–129 (1987).
14. E. Cassan, S. Laval, S. Lardenois, and A. Koster, "On-chip optical interconnects with compact and low-loss light distribution in silicon-on-insulator rib waveguides," IEEE J. Sel. Top. Quantum Electron. **9**(2), 460–464 (2003).
15. M. J. R. Heck, J. F. Bauters, M. L. Davenport, D. T. Spencer, and J. E. Bowers, "Ultra-low loss waveguide platform and its integration with silicon photonics," Laser Photon. Rev. **8**(5), 667–686 (2014).
16. P. D. Trinh, S. Yegnanarayanan, F. Coppinger, and B. Jalali, "Silicon-on-insulator (SOI) phased-array wavelength multi/demultiplexer with extremely low-polarization sensitivity," IEEE Photonics Technol. Lett. **9**(7), 940–942 (1997).
17. Q. Xu, B. Schmidt, S. Pradhan, and M. Lipson, "Micrometre-scale silicon electro-optic modulator," Nature **435**(7040), 325–327 (2005).
18. M. Casalino, G. Coppola, R. M. De La Rue, and D. F. Logan, "State-of-the-art all-silicon sub-bandgap photodetectors at telecom and datacom wavelengths," Laser Photon. Rev. **10**(6), 895–921 (2016).
19. G. Li, J. Yao, H. Thacker, A. Mekis, X. Zheng, I. Shubin, Y. Luo, J. Lee, K. Raj, J. E. Cunningham, and A. V. Krishnamoorthy, "Ultralow-loss, high-density SOI optical waveguide routing for macrochip interconnects," Opt. Express **20**(11), 12035 (2012).
20. M. Sartison, O. Camacho Ibarra, I. Caltzidis, D. Reuter, and K. D. Jöns, "Scalable integration of quantum emitters into photonic integrated circuits," Mater. Quantum Technol. **2**(2), 023002 (2022).
21. J.-H. Kim, S. Aghaeimeibodi, J. Carolan, D. Englund, and E. Waks, "Hybrid integration methods for on-chip quantum photonics," Optica **7**(4), 291 (2020).
22. P. Holewa, A. Sakanas, U. M. Gür, P. Mrowiński, A. Huck, B.-Y. Wang, A. Musiał, K. Yvind, N.

Gregersen, M. Syperek, and E. Semenova, "Bright Quantum Dot Single-Photon Emitters at Telecom Bands Heterogeneously Integrated on Si," ACS Photonics **9**(7), 2273–2279 (2022).
23. B. R. Bennett, R. A. Soref, and J. A. Del Alamo, "Carrier-induced change in refractive index of InP, GaAs and InGaAsP," IEEE J. Quantum Electron. **26**(1), 113–122 (1990).
24. M. I. Davanço, J. Liu, L. Sapienza, C. Z. Zhang, J. V. De Miranda Cardoso, V. Verma, R. P. Mirin, S. W. Nam, L. Liu, and K. Srinivasan, "Heterogeneous integration for on-chip quantum photonic circuits with single quantum dot devices," Nat. Commun. **8**, 889 (2017).
25. M. Zieliński, Y. Don, and D. Gershoni, "Atomistic theory of dark excitons in self-assembled quantum dots of reduced symmetry," Phys. Rev. B **91**, 085403 (2015).
26. P. Schnauber, A. Singh, J. Schall, S. I. Park, J. D. Song, S. Rodt, K. Srinivasan, S. Reitzenstein, and M. I. Davanço, "Indistinguishable Photons from Deterministically Integrated Single Quantum Dots in Heterogeneous GaAs/Si3N4 Quantum Photonic Circuits," Nano Lett. **19**, 7164–7172 (2019).
27. V. M. Ustinov, N. A. Maleev, A. E. Zhukov, A. R. Kovsh, A. Y. Egorov, A. V. Lunev, B. V. Volovik, I. L. Krestnikov, Y. G. Musikhin, N. A. Bert, P. S. Kop'ev, Z. I. Alferov, N. N. Ledentsov, and D. Bimberg, "InAs/InGaAs quantum dot structures on GaAs substrates emitting at 1.3 μm," Appl. Phys. Lett. **74**(19), 2815 (1999).
28. Ł. Dusanowski, P. Holewa, A. Maryński, A. J. Musiał, T. Heuser, N. Srocka, D. Quandt, A. Strittmatter, S. Rodt, J. Misiewicz, S. Reitzenstein, and G. Sęk, "Triggered high-purity telecom-wavelength single-photon generation from p-shell-driven InGaAs/GaAs quantum dot," Opt. Express **25**(25), 31122 (2017).
29. N. N. Ledentsov, A. R. Kovsh, A. E. Zhukov, N. A. Maleev, S. S. Mikhrin, A. P. Vasil'ev, E. Semenova, M. V Maximov, Y. M. Shernyakov, N. V Kryzhanovskaya, V. M. Ustinov, and D. Bimberg, "High performance quantum dot lasers on GaAs substrates operating in 1.5 μm range," Electron. Lett. **39**(15), 1126–1128 (2003).
30. E. S. Semenova, R. Hostein, G. Patriarche, O. Mauguin, L. Largeau, I. Robert-Philip, A. Beveratos, and A. Lemaître, "Metamorphic approach to single quantum dot emission at 1.55μm on GaAs substrate," J. Appl. Phys. **103**(10), 103533 (2008).
31. S. L. Portalupi, M. Jetter, and P. Michler, "InAs quantum dots grown on metamorphic buffers as non-classical light sources at telecom C-band: a review," Semicond. Sci. Technol. **34**(5), 053001 (2019).
32. A. J. Musiał, P. Holewa, P. Wyborski, M. Syperek, A. Kors, J. P. Reithmaier, G. Sęk, and M. Benyoucef, "High-Purity Triggered Single-Photon Emission from Symmetric Single InAs/InP Quantum Dots around the Telecom C-Band Window," Adv. Quantum Technol. **3**(2), 1900082 (2020).
33. Ł. Dusanowski, M. Syperek, J. Misiewicz, A. Somers, S. Höfling, M. Kamp, J. P. Reithmaier, and G. Sęk, "Single-photon emission of InAs/InP quantum dashes at 1.55 μm and temperatures up to 80 K," Appl. Phys. Lett. **108**(16), 163108 (2016).
34. A. Kors, K. Fuchs, M. Yacob, J. P. Reithmaier, and M. Benyoucef, "Telecom wavelength emitting single quantum dots coupled to InP-based photonic crystal microcavities," Appl. Phys. Lett. **110**(3), 031101 (2017).
35. P. Holewa, S. Kadkhodazadeh, M. Gawełczyk, P. Baluta, A. J. Musiał, V. G. Dubrovskii, M. Syperek, and E. Semenova, "Droplet epitaxy symmetric InAs/InP quantum dots for quantum emission in the third telecom window: morphology, optical and electronic properties," Nanophotonics (2022).
36. Y. A. Vlasov and S. J. McNab, "Losses in single-mode silicon-on-insulator strip waveguides and bends," Opt. Express **12**(8), 1622 (2004).
37. Http://www.lumerical.com/tcad-products/fdtd/, "Lumerical Solutions, Inc.," (n.d.).
38. P. Mrowiński, M. Emmerling, C. Schneider, J. P. Reithmaier, J. Misiewicz, S. Höfling, and G. Sęk, "Photonic engineering of highly linearly polarized quantum dot emission at telecommunication wavelengths," Phys. Rev. B **97**(16), 165427 (2018).
39. P. Mrowiński and G. Sęk, "Modelling the enhancement of spectrally broadband extraction efficiency of emission from single InAs/InP quantum dots at telecommunication wavelengths," Phys. B Condens. Matter **562**(March), 141–147 (2019).
40. A. Ghimire, E. Shafran, and J. M. Gerton, "Using a sharp metal tip to control the polarization and direction of emission from a quantum dot.," Sci. Rep. **4**, 6456 (2014).
41. L. Yang, H. Wang, Y. Fang, and Z. Li, "Polarization State of Light Scattered from Quantum Plasmonic Dimer Antennas.," ACS Nano **10**(1), 1580–8 (2016).
42. K. Żołnacz, A. J. Musiał, N. Srocka, J. Große, M. J. Schlösinger, P.-I. Schneider, O. Kravets, M. Mikulicz, J. Olszewski, K. Poturaj, G. Wójcik, P. Mergo, K. Dybka, M. Dyrkacz, M. Dłubek, S. Rodt, S. Burger, L. Zschiedrich, G. Sęk, S. Reitzenstein, and W. Urbańczyk, "Method for direct coupling of a semiconductor quantum dot to an optical fiber for single-photon source applications," Opt. Express **27**(19), 26772 (2019).
43. A. Dousse, L. Lanco, J. Suffczyński, E. Semenova, A. Miard, A. Lemaître, I. Sagnes, C. Roblin, J. Bloch, and P. Senellart, "Controlled light-matter coupling for a single quantum dot embedded in a pillar microcavity using far-field optical lithography," Phys. Rev. Lett. **101**, 267404 (2008).
44. M. Gschrey, F. Gericke, A. Schüßler, R. Schmidt, J.-H. Schulze, T. Heindel, S. Rodt, A. Strittmatter, and S. Reitzenstein, "In situ electron-beam lithography of deterministic single-quantum-dot mesa-structures using low-temperature cathodoluminescence spectroscopy," Appl. Phys. Lett. **102**, 251113 (2013).
45. J. Lee, I. Karnadi, J. T. Kim, Y.-H. Lee, and M.-K. Kim, "Printed Nanolaser on Silicon," ACS Photonics **4**(9), 2117–2123 (2017).

46. R. Katsumi, Y. Ota, M. Kakuda, S. Iwamoto, and Y. Arakawa, "Transfer-printed single-photon sources coupled to wire waveguides," Optica **5**(6), 691 (2018).
47. J.-H. Kim, S. Aghaeimeibodi, C. J. K. Richardson, R. P. Leavitt, D. Englund, and E. Waks, "Hybrid Integration of Solid-State Quantum Emitters on a Silicon Photonic Chip," Nano Lett. **17**(12), 7394–7400 (2017).
48. A. Sakanas, E. Semenova, L. Ottaviano, J. Mørk, and K. Yvind, "Comparison of processing-induced deformations of InP bonded to Si determined by e-beam metrology: Direct vs. adhesive bonding," Microelectron. Eng. **214**, 93–99 (2019).
49. G. Muliuk, N. Ye, J. Zhang, A. Abbasi, A. Trindade, C. Bower, D. Van Thourhout, and G. Roelkens, "Transfer Print Integration of 40Gbps Germanium Photodiodes onto Silicon Photonic ICs," in *2017 European Conference on Optical Communication (ECOC)* (IEEE, 2017), pp. 1–3.
50. G. Roelkens, L. Liu, D. Liang, R. Jones, A. Fang, B. Koch, and J. Bowers, "III-V/silicon photonics for on-chip and intra-chip optical interconnects," Laser Photon. Rev. **4**(6), 751–779 (2010).
51. Y. Fu, T. Ye, W. Tang, and T. Chu, "Efficient adiabatic silicon-on-insulator waveguide taper," Photonics Res. **2**(3), A41 (2014).
52. M. I. Davanço and K. Srinivasan, "Efficient spectroscopy of single embedded emitters using optical fiber taper waveguides," Opt. Express **17**(13), 10542 (2009).
53. S. Hepp, S. Bauer, F. Hornung, M. Schwartz, S. L. Portalupi, M. Jetter, and P. Michler, "Bragg grating cavities embedded into nano-photonic waveguides for Purcell enhanced quantum dot emission," Opt. Express **26**(23), 30614 (2018).
54. P. Mrowiński, P. Schnauber, P. Gutsche, A. Kaganskiy, J. Schall, S. Burger, S. Rodt, and S. Reitzenstein, "Directional Emission of a Deterministically Fabricated Quantum Dot-Bragg Reflection Multimode Waveguide System," ACS Photonics **6**, 2231–2237 (2019).
55. A. Mohanty, M. Zhang, A. Dutt, S. Ramelow, P. Nussenzveig, and M. Lipson, "Quantum interference between transverse spatial waveguide modes," Nat. Commun. **8**(1), 14010 (2017).

# Supplementary: Optimization of heterogeneously integrated InP-Si on-chip photonic components

PAWEŁ MROWIŃSKI,[1*] PAWEŁ HOLEWA,[1] AURIMAS SAKANAS,[2] GRZEGORZ SĘK,[1] ELIZAVETA SEMENOVA,[2] AND MARCIN SYPEREK[1]

[1]Department of Experimental Physics, Faculty of Fundamental Problems of Technology, Wrocław University of Science and Technology, Wybrzeże Wyspiańskiego 27, 50-370 Wrocław, Poland
[2]DTU Electro, Technical University of Denmark, Kongens Lyngby 2800, Denmark
[3]NanoPhoton-Center for Nanophotonics, Technical University of Denmark, 2800 Kongens Lyngby, Denmark
*pawel.mrowinski@pwr.edu.pl

## I. Evaluation of InP thickness dependent on-chip efficiency

In the section "method and model structure" of the manuscript, we commented on the height of the InP part of the InP/Si WG fixed at 580 nm, which is used in optimization studies. Now, we would like to test the choice of the InP height by performing simulations for the best taper structure (#3 in Fig. 3(b)). In Fig. S1, we demonstrate coupling efficiency for the first element of the device, which is related to the QD-confined dipole coupled to the modes of the hybrid InP/Si WG. The coupling efficiency is plotted as a function of the InP height ranging from 250 nm to 800 nm. Additionally, the plot contains dipole emission coupling efficiency to the fundamental $TE_1$ mode, to both $TE_1$ and $TE_2$ modes ($TE_1+TE_2$), and includes the contribution of hybridized mode HE ($TE1+TE2+HE$) that has a TE fraction in the range of 50-90%. The HE modes significantly contribute to effective coupling efficiency in the InP height range from 550 nm to 600 nm. The mode distribution profiles that imprint mainly on coupling efficiency are presented to the right of Fig. S1, plotted for the 580-nm-thick InP waveguide. In this case, the total coupling efficiency equals roughly 60.2%, as stated in the manuscript, however, at 550 nm, the coupling efficiency is slightly higher, i.e. 62.7%.

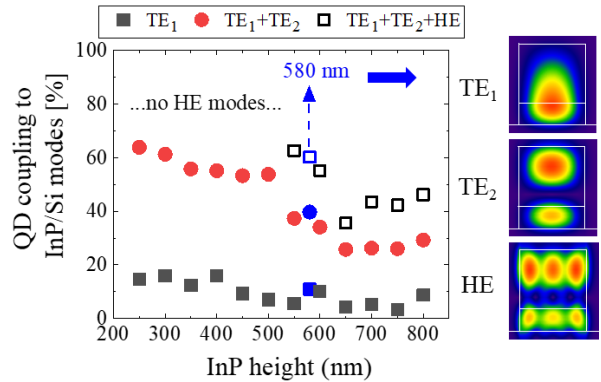

Figure S1. Calculated dipole emission (QD) coupling to the modes of the hybrid InP/Si waveguide with dependence on the InP height.

Next, we studied the optical field transmission from the hybrid InP/Si waveguide to the Si WG for the mentioned taper structure (#3 in Fig. 3(b)). Here, we use a similar approach to that shown in the manuscript relying on a propagating Gaussian source centered at the InP part of the hybrid waveguide. The results are demonstrated in Fig. S2a, showing transmission efficiency vs. the InP height. In the presented data one can observe a high transmission level (>

80%) in the whole range of InP heights, with the maximum at 600 nm (92.3%). A slightly lower transmission efficiency is found at 580 nm (86.4%). Lastly, we evaluated vertical outcoupling efficiency as a function of InP height for the outcoupler with the grating period R = [$\lambda_0 \cdot (3/4)/n_{eff}$ + $\lambda_0/4$] x 1.925, which resulted in the highest outcoupling efficiency towards the top direction at 1.55 µm photon wavelength within the numerical aperture of 0.65. The results are summarized in Fig. S2b). We observe a non-trivial dependence with the local maximum at 350 nm (20.0%) and the global maximum at 600 nm (25.4%). Here, the highest outcoupling efficiency is achieved for 580 nm-thick InP (25.6%).

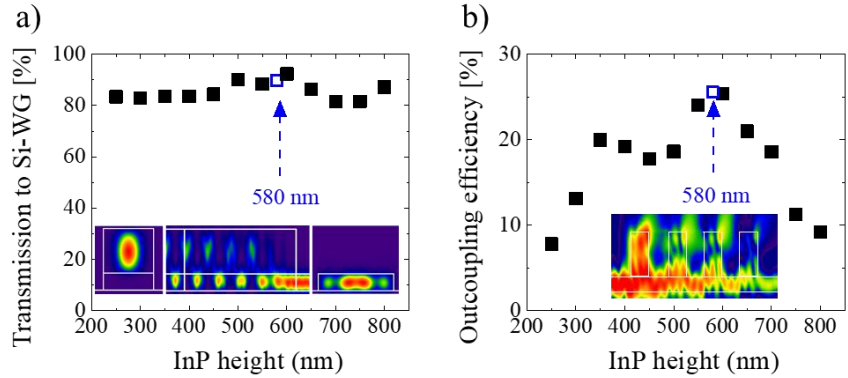

Figure S2. (a) Calculated transmission from the hybrid InP/Si waveguide to the Si waveguide via the taper structure and (b) vertical outcoupling efficiency via the circular Bragg grating structure, both in dependence on the InP height.

Finally, Fig. S3 summarizes obtained on-chip coupling efficiency of QD-confined dipole emission to the Si waveguide and vertical outcoupling efficiency using the optimized grating structure, both evaluated in dependence of the InP height. We obtained 53.3% on-chip efficiency with 550 nm-thick InP and a slightly lower efficiency of 52.0% with the 580 nm InP height. In the case of vertical outcoupling efficiency, we see that the 580 nm InP height is the optimum value for the highest outcoupling, reaching 13.8%. According to these results, we see that: i) the presence of the HE mode in the hybrid InP/Si waveguide improves QD-confined dipole emission coupling to the propagating field, ii) the optimized taper structure is rather insensitive to the change in the InP height, iii) the optimized vertical outcoupler for 580 nm reveals the maximum efficiency value. One can mention that the InP height set in the range of 200-400 nm would still be good enough for the efficient dipole coupling even without the contribution of HE modes, and for the efficient taper transmission, however, it would require an additional optimization step for the outcoupling design to improve the vertical outcoupling efficiency. We expect that more grating periods would be required to obtain a similar efficiency as for 580 nm, because the light transfer from Si to InP with a smaller height would be less effective. On the other hand, when the InP height is set above 600 nm, we expect a noticeable decrease in both the on-chip coupling and vertical outcoupling efficiency, which is related mainly to the less effective dipole coupling to a more complex multimode structure of the hybrid InP/Si WG. In this case, the solution to increase on-chip coupling efficiency would be to reduce the width of both InP and Si in the hybrid WG structure. Nonetheless, these changes would increase the risk that such a narrow and high system of WGs will be mechanically unstable, or possibly the excess losses of propagating modes would increase, either of which are not beneficial for the scalable photonic integrated circuits.

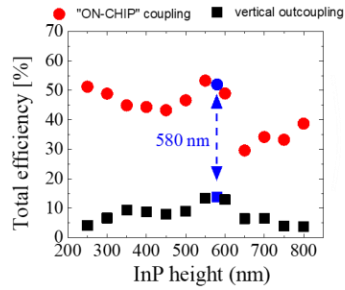

Figure S3. Evaluated on-chip and outcoupled efficiencies of the dipole emission in the hybrid InP/Si waveguide in dependence on the InP height.

II. **Evaluation of excess loss vs InP width**

This additional material demonstrates an extended analysis of the effective refractive index ($n_{eff}$) of modes confined in the hybrid InP/Si waveguides (WGs) in dependence of the width of the InP waveguide. Figure S4 demonstrates the field distribution of the first two TE-like modes (#1, and #3). It shows that in the case of the fundamental mode (#1), the optical field is more effectively transferred to InP with increasing WG width, while for the #3 mode, the field is distributed partially in the InP WG and partially in the Si WG. These results can be explained through the non-trivial dependence of the effective refractive index with the InP WG width, as is shown in Fig. S5. Here, in both modes, real part of the effective refractive index increases with increasing InP width. However, for the fundamental mode (#1), the excess losses related to the leakage along the WG increase as well, whereas the excess losses decrease with the #3 mode. The opposite behaviour for these modes is related to the noticeable transfer of the mode energy from Si to InP for the fundamental mode, which is not the case for the #3 mode, as visible in Fig. S5 on the very right panel.

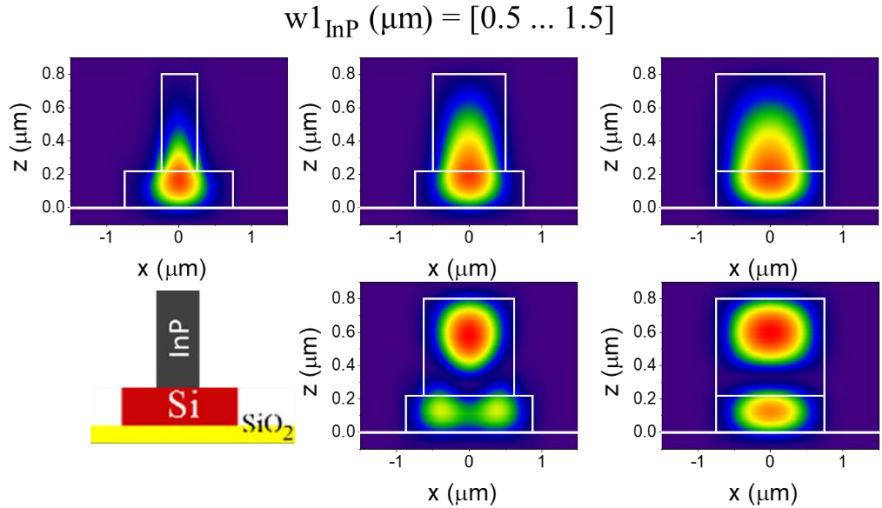

Figure S4. Electric field distribution of the TE-like fundamental mode (#1) and the higher order TE-like mode (#3) for the hybrid InP/Si waveguides in dependence on the InP waveguide width. A scheme of the hybrid InP/Si WG is shown in the bottom row on the left.

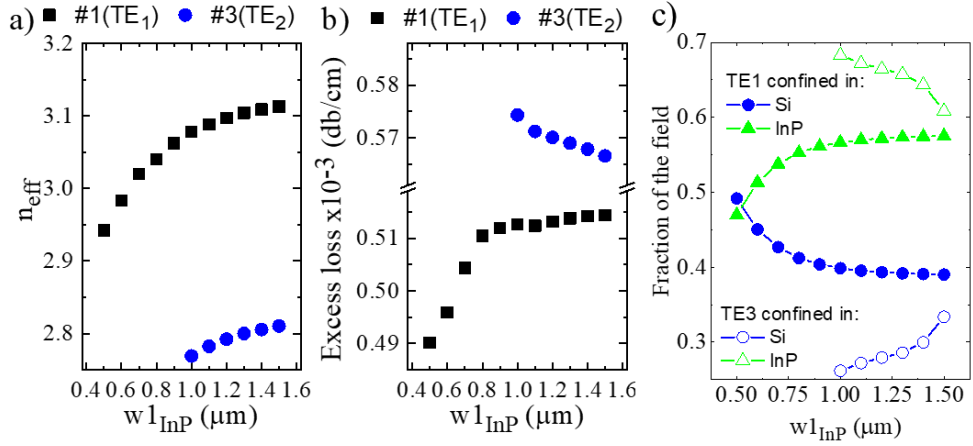

Figure S5. (a) The effective refractive index dependence on the InP waveguide width, together with (b) the losses evaluated for modes #1 and #3. The opposite trends in (b) are related to the transfer of the confined optical field from the Si WG to the InP WG, shown in (c).

### III. Settings of FDTD lumerical software used in sec. 3.1-3.3

A primary benefit of using the FDTD approach for photonic systems of our interest containing QDs as light sources, which can hardly be controlled in terms of emission wavelength, is a broadband solution obtained in a single simulation. However, the large size of the computational domain is a drawback which extends the simulation time, so the mesh settings must be verified first to achieve accurate results within a minute-time scale. Having a mesh cell of around 40x40x40 nm^3, we did not observe any stair-casing errors like atypical diffraction of the field propagating along the taper at 1.55 µm wavelength – roughly, it results in 10 mesh cells per wavelength in InP. The simulation time of each taper takes 5-10 min, where we used an auto-shutoff feature that stops the simulations when the energy level in the simulation region falls below 10^-5. The computational hardware is a standard personal computer with 20GB RAM and the i5-6300HQ Intel processor. Other important FDTD parameters are: default stretched coordinate PML (based on the formulation proposed by Gedney and Zhao [S. D. Gedney and B. Zhao, An Auxiliary Differential Equation Formulation for the Complex-Frequency Shifted PML, IEEE Trans. on Antennas & Propagat., vol. 58, no. 3, 2010.] as boundary conditions – 8 layers, simulation time 1000 fs, dt = 0.0779186 fs. In addition we must search the parameter space for the taper optimization on a stepwise approach, so we first did the quick scan of the taper geometry for a fixed length of the taper of 20 µm, then we performed a scan including additional iteration with respect to the taper length from 20 µm to 25 µm, and last we investigated the finest parameter space for the taper width w2InP and w2Si.

### IV. Evaluation of propagating modes

The propagating modes in a waveguide are calculated using a mode solver incorporated in Lumerical FDTD software. The output is for 1.55 µm wavelength, including the spatial mode profiles and effective index (real and imaginary part). The excess loss, which is related to the leakage along the waveguide, is calculated by

$$loss\ [db/m] = -10 log_{10}\left(\frac{P(z)|_{z=1m}}{P(z)|_{z=0m}}\right)$$

P(y) is the power of the field in the y position.

## V. Evaluation of taper transmission using fine scan.

In this section we demonstrate the evaluation of optical field transmission to the Si WG via taper structures by using a fine detuning of w2_Si and w2_InP parameters. For that purpose we apply position detuning to four tapers of exactly known geometries. All the cases revealed transmission above 85 %. Detuning for both w2_InP and w2_Si is performed by ±0.05 μm, ±0.10 μm, ±0.15 μm. In Fig. S6 we demonstrate a 2D representation of evaluated results. Firstly, we observe rather smooth fluctuations of the optical field transmission versus both detuning, within a range of ±3 %, showing that all geometries found in the parameter space demonstrated in Table 1 are close to the optimal value, as well as the tolerance for the fabrication processing of the order 50 nm is required and can be fulfilled by a processing method based on electron-beam lithography.

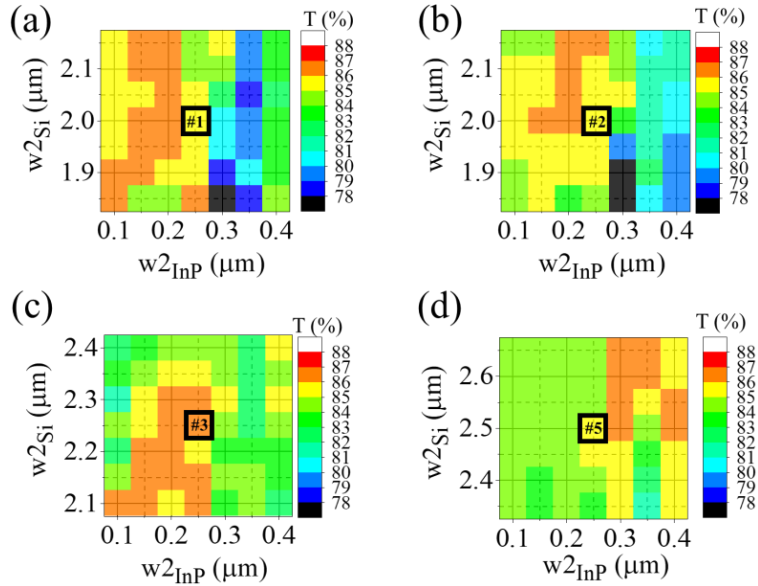

Figure S6. (a-d) Evaluation of taper transmission for fine detuning of w2InP and w2Si parameters.

## VI. Evaluation of dipole coupling to individual modes

A more detailed analysis of dipole coupling to individual modes in hybrid InP/Si waveguide is performed in dependence on the vertical displacement $\Delta z$ and horizontal displacement $\Delta x$. The results are demonstrated in Fig. S7. In Fig. S7(a) we demonstrate the individual coupling coefficients to the selected modes for a $\Delta z$ changing from -0.1 μm to 0.1 μm. We observe that coupling to higher order modes, i.e. #3 and #6 increases, while the coupling to fundamental mode, #1, is less effective, which in total results in increase of the overall coupling efficiency for the dipole shifted by $\Delta z = 0.1$ μm of 42.3 %. Next, in Fig. S7(b) we vary $\Delta x$ from 0 to 0.4 μm and we observe for $\Delta x = 0.25$ μm a significantly reduced coupling to a HE mode (#6), and reduced coupling to both fundamental mode, #1, and mode #3, although, it is compensated by activated coupling of about 10% to the mode #4, which finally reuslts in reduced coupling from

38 % to 27.3 %. A further shift along the 'x' axis to Δx = 0.4 μm results in still high overall coupling of 37.5 % due to a restored coupling to HE mode (#6) and still significant contribution of the coupling to mode #4.

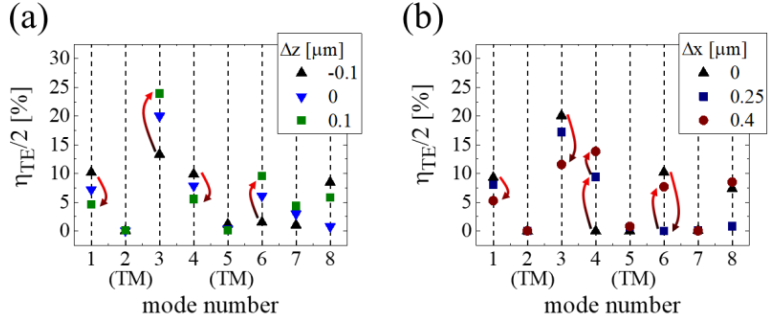

Figure S7. Dipole emission coupling to InP/Si waveguide modes for specific modes shown for (a) dipole displacements along (a) vertical 'z' axis and (b) horizontal 'x' axis.

VII. **Evaluation of mode matching from Si waveguide to InP/Si outcoupler**

In this section we demonstrate quick evaluation of the modes matching at the intersection of the Si WG (Fig. S8a) and InP ring structure on Si slab waveguide (Fig. S8b), which serves for outcoupling of light field towards external detection system. This is important to note that in our conceptual system the, due to refractive index contrast between the InP and Si ($n_{InP} < n_{Si}$), the mode matching between the fundamental modes (Fig. 8(c,d)), which is given by:

$$\eta_m = \frac{|\int E_1^* E_2 dA|^2}{\int |E_1|^2 dA \int |E_2|^2 dA},$$

is relatively high (~60 %). However, we note that coupling of fundamental mode of Si WG with higher order modes of the outcoupler section is low (<1 %).

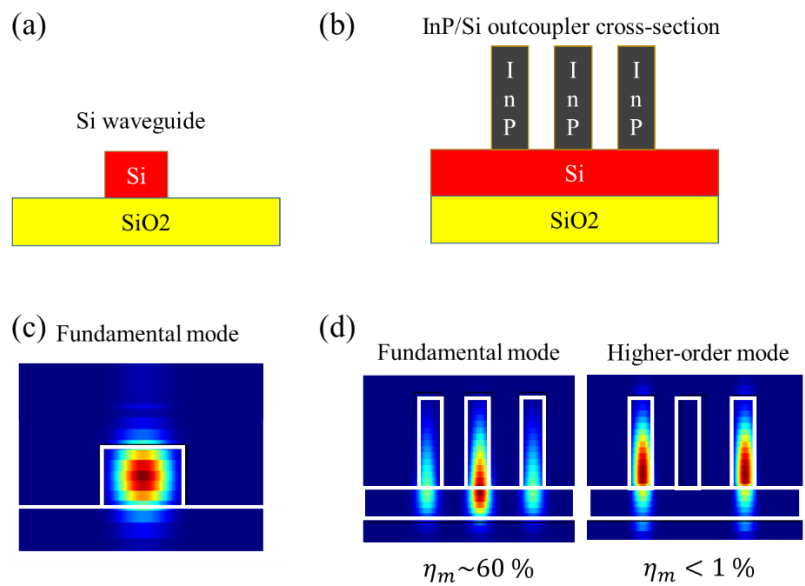

Figure S8. (a) Schematic view of Si waveguide (WG) cross-section and (b) of InP ring structure cross-section, which are used for evaluation of mode matching between modes. (c) A field profile of a fundamental mode in Si WG and (d) field profiles of fundamental and higher-order modes in InP ring structure. The values of $\eta_m$ are evaluated for the matching with the mode shown in (c).